\documentclass[twocolumn,showpacs,preprintnumbers,amsmath,amssymb,superscriptaddress]{revtex4}
\usepackage{graphicx}% Include figure files
\usepackage{dcolumn}% Align table columns on decimal point
\usepackage{bm,color}% bold math
\begin{document}

%\preprint{APS/123-QED}
\title{Short-time motion of Brownian particles in a shear flow\\
}
% Force line breaks with \\

\author{Takuya Iwashita}
 \email{iwashita@cheme.kyoto-u.ac.jp}
\affiliation{Department of Chemical Engineering, Kyoto University, Kyoto 615-8510, Japan}
\author{Ryoichi Yamamoto}
 \email{ryoichi@cheme.kyoto-u.ac.jp}
\affiliation{Department of Chemical Engineering, Kyoto University, Kyoto 615-8510, Japan}
\affiliation{CREST, Japan Sience and Technology Agency, Kawaguchi 332-0012, Japan}
\date{\today}
%y}% It is always \today, today,
\begin{abstract}
The short-time motion of Brownian particles in an  incompressible Newtonian fluid under shear, in which the fluid inertia becomes important, was investigated by direct numerical simulation of particulate flows. Three-dimensional simulations were performed, wherein  external forces were introduced to approximately form Couette flows throughout the entire system with periodic boundary conditions. In order to examine the validity of the method, the mean square displacement of a single spherical particle in a simple shear flow was calculated, and these results were compared with a hydrodynamic analytical solution that includes the effects of the fluid inertia. Finally, the dynamical behavior of a monodisperse dispersion composed of repulsive spherical particles was examined on short time scales, and the shear-induced diffusion coefficients were measured 
for several volume fractions up to 0.50.
\end{abstract}

%82.20.Wt Computational modeling; simulation
%82.70.-y Disperse systems; complex fluids
%05.40.-a Fluctuation phenomena, random processes, noise, and Brownian motion
%83.50.Ax Steady shear flows, viscometric flow 
\maketitle
\section{Introduction}
%About short-time motions
The short-time motion of small particles fluctuating in a Newtonian fluid is strongly affected by fluid inertia, that is, the vorticity of the host fluid surrounding the dispersed particles. 
If a dispersed particle accelerates due to Brownian forces, it affects the motion of the host fluid in the neighborhood of the particle, while the vorticity generated by the particle's motion then affects the motion of the same particle. 
These effects are referred to as memory effects and have an important role in the dynamical motion of a dispersion on short time scales \cite{Weitz,Zhu,lukic}. 

%Time scales and gorvening equation.
The vorticity diffuses away on a kinematic time scale $\tau_\nu = a^2/\nu$, where $a$ is the radius of the  spherical particle and $\nu$ is the kinematic viscosity of the host fluid ($\sim 10^{-6}$ s for $1\ \mu$m in water). 
When the vorticity diffuses away much faster than the  particle's motion, {\it i}.{\it e}. $\tau_\nu \ll \tau_B=M/6\pi\eta a$ where $M$ is the mass of a Brownian particle and $\eta$ is the shear viscosity of the host fluid, the motion of a Brownian particle is well approximated by the normal Langevin equation,  which is valid for strong damping (Reynolds number $Re\rightarrow 0$) or long time scales; 
however, 
this equation, wherein the effects of the fluid inertia are ignored is not applicable to a dispersion composed of neutrally buoyant particles since $\tau_B$ is comparable to $\tau_\nu$. 
For a complete understanding of the short-time motion of a dispersion, the inertias of the particle and the host fluid cannot be neglected. 
%

% numerical approach
One way to account for memory effects is to simultaneously resolve the fluid motion with the particle motion as a boundary condition to be satisfied.
We refer to this approach as the direct numerical simulation(DNS) approach. 
Within the DNS approach, various numerical methods have been developed \cite{LB,FP,LB0,SR,NS,IBM}, and the power-law decay behavior in the velocity autocorrelations of a free Brownian particle has been accurately reproduced. 
This slow relaxation of the correlation behavior is one of the main features of memory effects. 
Although most of these methods have been applied to  dispersions composed of free Brownian particles at thermal equilibrium, dispersions under flows that are far from equilibrium have not been examined in detail, even for the simple case of a single Brownian particle in a shear flow on a short time scale, $\tau_{\nu}$. 
Furthermore, most numerical methods used for concentrated dispersions under shear, which are widely used for measuring rheological properties, are limited to a Reynolds number of zero, and the short-time motions of the dispersed particles in concentrated dispersions cannot be correctly tracked.

Recently, we have developed a numerical method, known as the "Smoothed Profile Method (SPM)" \cite{key1,key2}, for the DNS of particulate flows.
Its computational accuracy and efficiency have been examined carefully
by Lio {et al.} for several flow problems \cite{err}.
The SPM has been applied to a dispersion composed of free Brownian particles at thermal equilibrium \cite{key3,key5}.
We have also succeeded in reproducing the power-law decay behavior in the translational and rotational velocity autocorrelations of a Brownian particle, and these results   
agree well with hydrodynamic analytical solutions for a free Brownian particle in an infinite fluid that accounts for memory effects.
%
%The slow relaxations of a free Brownian particle are also verified experimentally on the kinematic time-scales $\tau_\nu$\cite{Weitz, lukic}.
%

In order to simulate dispersions in non-equilibrium conditions on short time scales in which memory effects become significant, we have modified the SPM.
The primary objective of the present work is to accurately examine the short-time motion of Brownian particles in a simple shear flow by using the modified SPM.
In this paper, we first present the modified SPM, in which external forces are introduced to impose a shear flow into the system. 
In order to validate the method, we next compare the numerical results for the mean square displacement (MSD) with a hydrodynamic analytical solution of a Brownian particle in simple shear flows that account for fluid inertia. 
Furthermore, we apply our method to a dispersion composed of many spherical particles under shear. 
The MSD in the vorticity direction is then calculated for several volume fractions, and the time evolution is discussed.
Finally, the shear-induced diffusion coefficients are measured from the long-time behavior of the MSD, and the dependence of the determined diffusion coefficients on the volume fraction is examined.   

\section{Simulation Method}
Let us consider a monodisperse dispersion of repulsive spherical particles in a Newtonian host fluid. 
The dispersion is subjected to shear by an external force. 
The position of the $i$th particle is ${\bm R_i}$, the translational velocity is  ${\bm V_i}$, and the rotational velocity is ${\bm \Omega_i}$. 
The velocity and pressure fields of the host fluid are $\bm v(\bm x, t)$ and $p(\bm x, t)$, respectively. 
These field quantities are defined on three-dimensional Cartesian grids; $\bm x \in [0,L_x] [-L_y/2,L_y/2][0,L_z]$. 
In order to distinguish the particle and fluid domains on the grids, a smoothed function $\phi(\bm x, t)$,  which is equal to $1$ in the particle domains and $0$ in the fluid domains, is introduced. 
These domains are separated by a thin interfacial domain of thickness $\xi$. 
%
%dimension  
The length unit is taken to be the lattice spacing $\Delta$, and the time unit is $\rho_f\Delta^2/\eta$,  where $\rho_f$ denotes the density of the host fluid.
The time evolution of the $\it i$th dispersed particle with mass $M_i$ and moment of inertia $\bm I_i$ is governed by Newton's equations of motion:
\begin{align}
M_i \dot {\bm V_i}&= {\bm F^H_i} + {\bm F^C_i} + {\bm G_i^V},\ \ \
\dot {\bm R_i} = {\bm V_i},\\
{\bm I_i}\cdot \dot{\bm \Omega_i} &= {\bm N^H_i} + {\bm G_i^\Omega},
\end{align}
where $\bm F^H_i$ and $\bm N^H_i$ are the hydrodynamic forces and torques exerted by the host fluid on the particle \cite{key1,key2}.
$\bm F_i^C$ is a repulsive force that is employed to prevent particle overlaps, and a truncated Lennard-Jones potential, $V(r_{ij})=4[(\sigma/r_{ij})^{36}-(\sigma/r_{ij})^{18}+1/4]$ for $r_{ij}<2^{1/18}\sigma$  or $V(r_{ij})=0$, is adopted in this work. 
Here $r_{ij}=|\bm R_i - \bm R_j|$, and $\sigma=2a$ represents the diameter of particles.
%
%\textcolor{red}{For distances less than the lattice spacing, the lubrication force between the particles is not included, due to the space resolution.}
%
$\bm G_i^V$and $\bm G_i^\Omega$ are random forces and torques, respectively, due to thermal fluctuations. 
These fluctuations are introduced as white noise with a zero mean and correlations $ \langle \bm G_i^V(t) \bm G_j^V(0)\rangle= \alpha^V \bm I\delta(t)\delta_{ij}$ and $\langle \bm G_i^{\Omega}(t) \bm G_j^{\Omega}(0)\rangle= \alpha^\Omega \bm I\delta(t)\delta_{ij}$, where $\alpha^V$ and $\alpha^\Omega$ are numerical parameters that control the translational and rotational particle temperatures, namely, $T^V$ and $T^\Omega$ \cite{key3,key5}. 
The angular brackets denote taking an average over an equilibrium ensemble.
The temperatures are determined by the following procedure;
First a single Brownian particle at thermal equilibrium is simulated with fixed $\alpha_V$ and $\alpha_{\Omega}$. 
Then the translational and rotational long-time diffusion coefficients are obtained from the simulation. 
By comparing these diffusion coefficients with Stokes-Einstein diffusion coefficients, $D^V_0=k_BT^V/6\pi\eta a$ for the translational motion and $D^{\Omega}_0=k_BT^{\Omega}/8\pi\eta a^3$ for the rotational motion, we finally can determine the temperatures. 
Since both $\alpha_V$ and $\alpha_{\Omega}$ are chosen to satisfy $k_BT^V=k_BT^{\Omega}$, the temperatures are simply written as $k_BT$ throughout this paper.

\begin{figure}
\begin{center}
\includegraphics[scale=0.5]
{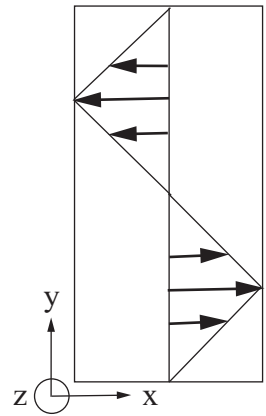}
\end{center}
\caption{\label{profile}{Schematic diagrams of a zigzag velocity flow. The arrows represent the velocity vectors of the host fluid in the flow direction. The $x$, $y$, and $z$ axis represent the flow, velocity-gradient, and vorticity directions, respectively}}
\end{figure}

The time evolution of the host fluid is governed by the Navier-Stokes equations:
\begin{align}
\rho_f(\partial_t {\bm v} + {\bm v}\cdot \nabla {\bm v}) &= \nabla \cdot\bm \sigma
 +\rho_f\phi{\bm f_p} +\rho_f \bm f^{shear},
\label{nseq}\\
\bm \sigma &= -p\bm I + \eta\{\nabla \bm v + (\nabla \bm v)^T\}
\end{align}
where the incompressibility condition, $\nabla \cdot \bm v=0$. $\bm f^{shear}(\bm x, t)$ is an external force field that is introduced to enforce the following velocity profile,
%in the system:
\begin{align}
v_{x}(y) &=
\begin{cases}
 \dot\gamma (-y - L_y/2), &(-L_y/2<y\leq -L_y/4)\\
 \dot\gamma y,  &(-L_y/4<y\leq L_y/4)\\
\dot\gamma (-y + L_y/2) &(L_y/4<y\leq L_y/2)\\
\end{cases}
\end{align}
where $\dot\gamma$ denotes the shear rate of the imposed flow and $y$ denotes the distance in the velocity-gradient direction.
This velocity profile, schematically depicted in Fig. \ref{profile}, enables us to solve the motion of the host fluid with periodic boundary conditions. 
Couette flow profiles are approximately formed over a range from $y=-L_y/4$ to $y=L_y/4$, although the zigzag profile strictly becomes a Couette flow when $L_y \rightarrow \infty$. 
$\phi {\bm f_p}$ represents the body force that ensures the rigidity of particles and the aprropriate non-slip boundary conditions at the fluid/particle interface, which is further elaborated in reference \cite{key1,key2, err}.

\section{Results and Discussion} 
The computational domain has three dimensions $64 \times 64 \times 64$ and periodic boundary conditions. 
The numerical parameters for both the host fluid and the spherical particles are $\Delta=1$, $\eta=1$, $\rho_f=1$, $a=5$, $\xi=2$, and particle density $\rho_p=1$.
The imposed shear rate is $\dot\gamma=0.005$, and the temperature is $k_BT=0.07$.
The temperature was determined by measuring the long-time diffusion coefficient of a single Brownian particle at 
thermal equilibrium before the shear was imposed.
This system has dimensionless parameters such that the Peclet number $Pe=6\pi\eta a^3\dot\gamma/k_BT\simeq 170$ and the particle Reynolds number $Re_p=\rho_p\dot\gamma a^2 /\eta= 0.125$. 
%
%This corresponds to a bouyant particle of the radius $a\sim 1.5\times 10^{-4}m$ in water at $300K$.
%
The initial configuration of the spherical particle is located at the central position of the system. 
The volume fraction of a single particle is $\Phi=0.002$.
%

%% Theoretical aspect
%We briefly explain theoretical aspects of a Brownian particle in simple shear flow.
%
An important feature of a single Brownian particle in a simple shear flow is that the MSD in the flow direction varies asymptotically with time as $t^3$. 
Although we would expect t-dependence of the MSD in the long-time limit, this non-diffusional behavior is explained as a result of a coupling between diffusive motion in the velocity-gradient direction and convective motion due to shear in the flow direction.
Theoretical solutions to this problem show the MSD in the flow direction($x$) is composed of three parts:
\begin{align}
\langle R^x_i(t)^2 \rangle = v^2_x(y_0)t^2 +\langle R^x_i(t)^2 \rangle_{\dot\gamma} + \langle R^x_i(t)^2 \rangle_0 \label{msdx} 
\end{align}
assuming that, at $t=0$, the particle is in the initial position ($0$, $y_0$, $0$),
wherein the first term on the right hand side of Eq. (\ref{msdx}) is a shear contribution only, and represents the simple translation of a particle along the streamline in the flow direction. 
The second term is the coupling term described above,  which represents the convection induced by diffusion. 
The third term, which is in the same form as the MSD of a free Brownian particle, is a thermal contribution only. 
Similar to that of the $x$ direction, the MSDs in the velocity-gradient direction($y$) and the vorticity direction($z$) are derived theoretically;
$\langle R^y_i(t)^2 \rangle=\langle R^y_i(t)^2 \rangle_0$ and 
$\langle R^z_i(t)^2 \rangle=\langle R^z_i(t)^2 \rangle_0$.  
These quantities are determined analytically for a single Brownian particle obeying the normal Langevin equation (LE) or the generalized Langevin equation (GLE) with memory effects \cite{ANA,ANA1,ANA2}.

\begin{figure}[hbt]
\begin{center}
\includegraphics[scale=0.8]
{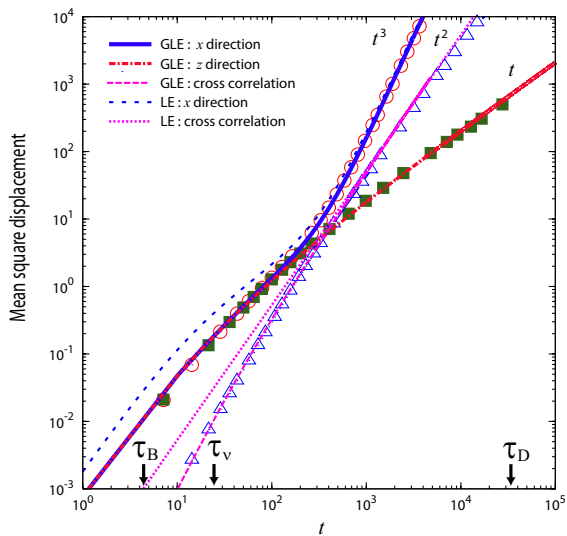}
\end{center}
\caption{\label{pict2}{Mean square displacement scaled by $k_BT$ of a single spherical particle fluctuating in a Newtonian fluid under shear at $k_BT=0.07$ and $\dot\gamma=0.005$: ($\bigcirc$) flow direction, ($\blacksquare$) vorticity direction, and ($\bigtriangleup$) positional cross-correlation.
The analytical solution of the MSD in the flow direction ($x$) for a Brownian particle in a shear flow: the dashed line represents the LE and the solid line the GLE. {The dash-dotted line is the analytical solution of the MSD in the vorticity direction ($z$) for the GLE, which is the same form as that for the GLE of a free Brownian particle}
The analytical solution of the positional cross-correlation for a Brownian particle in a shear flow: the dotted line represents the LE and the dashed line the GLE. The Brownian time $\tau_B\simeq 4.5$, the kinematic time $\tau_\nu=25$, and the diffusion time $\tau_D\simeq 3.4\times 10^4$. The parameters for both the particle and fluid are $a=5$, $\xi=2$, $\rho_p=1$, $\rho_f=1$, and $\eta=1$. The time unit is $\Delta^2/\nu=1$, and the length unit is $\Delta=1$.}}
\end{figure}

%% Numerical results :x direction
The time evolution of the MSD in the $x$ direction ($\bigcirc$) for a single spherical particle under shear is shown in Fig. \ref{pict2}. 
The MSD was calculated via 
$\langle |R_i^x(t)-R^x_i(0)-v_x(y_0)t|^2\rangle$
to eliminate the purely shear contribution that corresponds to the first term of Eq. (\ref{msdx}).
The MSD is scaled by $k_BT$.
The MSD is in excellent agreement with 
the analytical solution for the GLE.
For short times of up to $t\sim10^2$, the motion of the Brownian particle in the $x$ direction is like that of a free Brownian particle obeying the GLE. 
%
%Diffusive motion is not observed,
%while the diffusive region is approached for a Brownian particle obeying the LE on {\bf \underline{its time scales}}.
%
For long times, $t\gg \tau_\nu=25$, the MSD asymptotically approaches  $2D_0\dot\gamma^2 t^3/3$,  where $D_0=k_BT/6\pi\eta a$.
This $t^3$ regime in the MSD is approached in a much slower manner, $t^{-1/2}$, than that of the analytical solution for the LE.
The transient behavior of the MSD depends strongly on  whether the memory effects are taken into account. 

%% Numerical results: z direction
The MSD $\langle |R^z_i(t)-R^z_i(0)|^2\rangle$ in the $z$ direction ($\blacksquare$) is also plotted in Fig. \ref{pict2}. 
The numerical results of the MSD in the $z$ direction agree well with the analytical solution for the GLE of a free Brownian particle where the $t$ regime is approached in a much slower manner, $t^{-1/2}$, and the diffusive motion is attained on time scales of $O(10^4)$. 
The diffusion time characterizing the diffusive motion is $\tau_D=a^2/D_0\simeq 3.4 \times 10^4$, which measures the particle diffusion over the particle size.  
 
%%Numerical results: cross-correlation
Another dynamical quantity of interest is the positional cross-correlation $\langle R^x_i(t)R^y_i(t)\rangle$ of a Brownian particle in a simple shear flow.
This analytical form is derived for both the LE and GLE.
These long-time behaviors show the same dependence on $t$, which is $D_0\dot\gamma t^2$. 
The cross-correlation was calculated via $\langle (R_i^x(t) - R_i^x(0)-V^x_i(0)t)(R_i^y(t)-R_i^y(0))\rangle$.
The numerical results ($\bigtriangleup$) are plotted in Fig. \ref{pict2} and are in good agreement with the analytical solution for the GLE.
%

%%disccusion our model
This detailed analysis of the MSD is sufficient to confirm the validity of our method of incorporating  memory effects; 
however, within our method, it is assumed that the thermal fluctuations can be represented by white and gaussian noise. 
This means that correlations between thermal noise at different times are completely ignored, although  
in the theoretical framework of the GLE, noises are correlated.
The thermal noise memory, which we ignore, is considered to affect the particle's motion on very fast time scales $t<\tau_B\sim 4.5$. 
On these very short time scales, $t<\tau_B$, 
a gap between the hydrodynamic analytical solutions and the numerical results is clearly observed in  previous studies \cite{key3, key5}.
Although our method is not applicable to particle motion for $t<\tau_B$, the method can be accurately applied to the motion of particles in a shear flow on time scales $t>\tau_B$, in which the memory effects  become significant.

%% Abount concentrated dispersions 
Most numerical approaches so far performed for concentrated
particles dispersions are based on the Stokesian dynamics \cite{SD}, 
thus those simulations are valid only for the zero Reynolds number limit.
Recently, concentrated dispersions at finite Reynolds number have been 
simulated using the DNS methods based on the stochastic rotation dynamics \cite{SRDm} or the lattice Boltzmann method \cite{LB1, LB2, LB3},
We have applied the modified SPM to concentrated dispersions at a finite
Reynolds number $Re_p=0.125$ in order to examine the short-time motion 
of Brownian particles in a shear flow at finite volume fractions.
Simulations were performed at several volume fractions from a very dilute
case with $\Phi=0.002$ to very dense cases with $\Phi=0.4$ and $0.5$,
which are apparently higher than the previous DNS simulations.
%
%
%Concentrated dispersions are of great importance in the rheology of dispersions. 
%

Since $Pe\simeq170$, the hydrodynamic shear forces become dominant over the thermal forces in the particle' motion. 
The initial configurations of the dispersed particles are randomly distributed. 
%
%All simulations are performed in an isotropic disorder phase.
%

\begin{figure}[tbh]
\begin{center}
\includegraphics[scale=0.9]{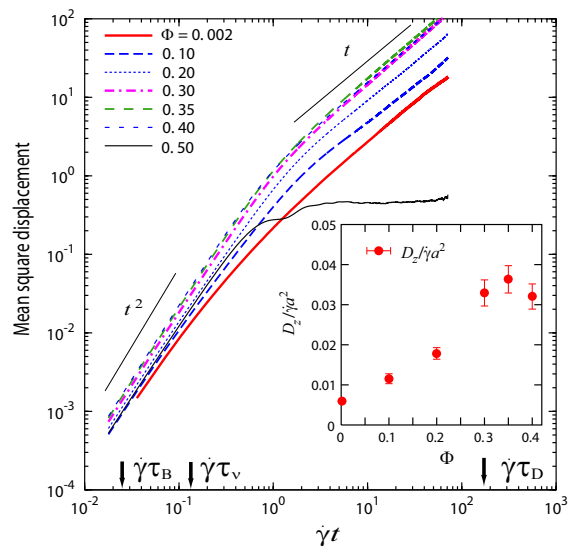}
\end{center}
\caption{\label{msd_t3}{The time evolution of the mean square displacement in the vorticity direction ($z$) for several volume fractions $\Phi$ at $k_BT=0.07$ and $\dot\gamma=0.005$. The Peclet number $Pe\simeq 170$ and the particle Reynolds number $Re_p=0.125$. 
%The solid line represents the hydrodynamic analytical solution of a Brownian particle subjected to shear at an infinite dilution. 
The inset shows the volume fraction dependence of $D_z$ scaled by $\dot\gamma a^2$. 
%The solid curve is a guide to the eye. %
The parameters for both the particle and fluid are $a=5$, $\xi=2$, $\rho_p=1$, $\rho_f=1$, and $\eta=1$.}}
\end{figure}

Figure \ref{msd_t3} shows the time evolution of the MSD in the $z$ direction for several volume fractions, up to 
%$\phi=0.4$. 
$\Phi=0.5$.
The time is scaled by $1/\dot\gamma$.
As the volume fraction is increased,
the MSD grows more rapidly in time due to the hydrodynamic and direct interactions between the particles, resulting in a increase in the slope of the MSD for a small strain $\dot\gamma t\sim O(10^{-1})$.
% 
%For $\phi\geq 0.2$, 
For high volume fractions $0.2 \leq \Phi \leq 0.4$,
the diffusive behavior is attained at a smaller strain $\dot\gamma t \sim O(1)$ rather than at $\Phi=0.002$, as for a Brownian particle.
The accelerated Stokesian dynamics (ASD) for non-Brownian particles at $\Phi=0.2$ show 
the diffusive region is attained at larger strains 
than at least 10 \cite{ASD}
%a transient region to a linear behavior that exists up to strains of at least 10 \cite{ASD}.
%
Compared with these numerical results, the onset of the diffusive region in the present simulations is much faster. 
The time at which the diffusive region is attained shifts to shorter times at higher volume fractions. 

For the highest volume fraction $\Phi=0.5$, however, we see that the motion of 
Brownian particles are trapped within the effective cages formed by the
surrounding particles.
This is because the particles start to form string-like objects in the
flow direction, and finally the whole system evolves into a
two-dimensional ordered structure in the plane perpendicular to the
flow,
similarly to the flow-induced ordering commonly observed in experiments \cite{EXOD1,EXOD0,EXOD2} and by computer simulations 
\cite{NMD,SD01,SD00} under shear flow.
In the present case with $Re_p = 0.125$ and $Pe \sim 170$,
we found that the shear-induced ordering occurs at the high volume fraction 
between $\Phi=0.4$ and $0.5$.

%%%%%%%%%%%%%%%%%%%%%
The long-time diffusion coefficient $D_{z}$ was calculated by linearly fitting the data over the diffusion regions.
The inset of Fig. \ref{msd_t3} shows the volume fraction dependence of $D_{z}$ scaled by $\dot\gamma a^2$. 
The diffusion coefficient increases rapidly up to $\Phi=0.3$ and reaches a plateau at a volume fraction beyond $\Phi>0.3$.
This behavior of $D_z$ is remarkably different from that at thermal equilibrium, where the diffusion coefficient decreases with increasing volume fraction. 
The enhancement of the diffusion coefficient with increasing volume fraction is a typical characteristic of shear-induced diffusion coefficients. 
These results exhibit the same qualitative behavior as the experimental results obtained for non-Brownian particles  
\cite{EXP0,EXP,EXP1}. 

\section{Conclusion}
In conclusion, the short-time motion of Brownian particles in an incompressible Newtonian fluid under shear was examined by using a modified SPM in which external forces are introduced to approximately form a simple shear flow throughout the entire system with periodic boundary conditions. 
The validity of the method was carefully examined by comparing the present numerical results for the MSD with the hydrodynamic analytical solution of the GLE of a single Brownian particle in a simple shear flow. 
{In the present study, we aimed to modify the original SPM by
incorporating thermal fluctuations so that the modified SPM is valid for
$\tau_B<t$, while other computational methods such as Brownian dynamics
and ASD, which are based on the LE that ignores memory effects due to
fluid motions, are valid only for $\tau_B\ll \tau_D<t$.}
Simulations were then performed for monodisperse dispersions of repulsive spherical particles at volume fractions ranging from $\Phi=0.002$-$0.50$.
We found that the MSD in the vorticity direction grows rapidly in time and with increasing particle volume fraction. At a strain $\dot\gamma t\sim O(1)$, the diffusive region is attained for $0.2 \leq \Phi \leq  0.4$.
The onset of the diffusive region shifts to shorter times at higher volume fractions. 
For $\Phi=0.5$, however, the particles are no more diffusive because of  
the shear-induced ordering.
The diffusion coefficient in the vorticity direction was obtained from the long-time behavior of the MSD. 
For volume fractions of up to $0.3$, the diffusion coefficient rises rapidly with increasing volume fraction. 
It then levels off for volume fractions beyond $0.3$. 
%
%A detailed analysis of the short-time motions in concentrated dispersions is in progress. 

%\bibliographystyle{junsort}
%\bibliography{brownian_theory,colloid_sim,brownian_exp,book,fluid%}

\end{document}